\documentclass[preprintnumbers,amsmath,amssymb]{revtex4}
\usepackage{amsmath} 
\usepackage{graphicx} 

\newcommand{\real}{{\mathbb R}}
\pdfoutput=1
\begin{document}

\title{Quantum measurements as weighted symmetry breaking processes: the hidden measurement perspective}

\author{Diederik Aerts}
\affiliation{Center Leo Apostel for Interdisciplinary Studies and Department of Mathematics, \\
Brussels Free University, Brussels, Belgium}\date{\today}
\email{diraerts@vub.ac.be} 
\author{Massimiliano Sassoli de Bianchi}
\affiliation{Laboratorio di Autoricerca di Base, 6914 Lugano, Switzerland}\date{\today}
\email{autoricerca@gmail.com} 

\date{\today}

\begin{abstract}

\noindent The purpose of the present note is twofold. Firstly, we highlight the similarities between the ontologies of Kastner's possibilist transactional interpretation (PTI) of quantum mechanics -- an extension of Cramer's transactional interpretation -- and the authors' hidden-measurement interpretation (HMI). Secondly, we observe that although a weighted symmetry breaking (WSB) process was proposed in the PTI, to explain the actualization of incipient transactions, no specific mechanism was actually provided to explain why the weights of such symmetry breaking are precisely those given by the Born rule. In other terms, PTI, similarly to decoherence theory, doesn't explain a quantum measurement in a complete way, but just the transition from a pure state to a fully reduced density matrix state. On the other hand, the recently derived extended Bloch representation (EBR) -- a specific implementation the HMI -- precisely provides such missing piece of explanation, i.e., a qualitative description of the WSB as a process of actualization of hidden measurement-interactions and, more importantly, a quantitative prediction of the values of the associated weights that is compatible with the Born rule of probabilistic assignment. Therefore, from the PTI viewpoint, the EBR provide the missing link for a complete description of a quantum measurement. However, EBR is in a sense more general than PTI, as it does not rely on the specific notion of transaction, and therefore remains compatible with other physical mechanisms that could be at the origin of the measurement-interactions.
\\

\noindent {\bf Keywords} Possibilist transactional interpretation $\cdot$ Hidden-measurement interpretation $\cdot$ Extended Bloch representation $\cdot$ Weighted symmetry breaking $\cdot$ Born rule $\cdot$ Non-spatiality

\end{abstract}

\maketitle

\section{Possibilist transactional interpretation}

Consider the following four fundamental assumptions: 

\begin{quote}

(1) A Hilbert state-vector describes the real condition and modes of being of a physical entity, and not our beliefs about them;

(2) The reduction of a state-vector during a measurement is an objective process of the change of state, and not a subjective process of the change of belief;

(3) The experimenter's consciousness does not play any causal role in a measurement process;

(4) A microscopic entity, like an electron, is a non-spatial entity which can only be spatialized when interacting with a measurement apparatus, or when incorporated into a macroscopic object.

\end{quote}

\noindent As far as we know, only two interpretations of quantum mechanics take the above four assumptions seriously: Kastner's possibilist transactional interpretation (PTI) \cite{k2012,k2013} and Aerts' Hidden-measurement interpretation (HMI) \cite{a1998,a1999,asdb2014}. Curiously, the initial versions of these two interpretations were “synchronously” reported for the first time in 1985 \cite{c1985,a1985} and both were subsequently published in a physics journal in 1986 \cite{c1986,a1986}. To say it all, the same synchronicity also holds for the so-called objective collapse theories, a first version of which was proposed by G. Ghirardi, A. Rimini and T. Weber \cite{grw1985,grw1986}. However, and as far as we know, the latter only agree with the first three of the above four assumptions. In any case, in the present article we shall only be interested in establishing a parallel between PTI and HMI, and more specifically to that implementation of the latter called the extended Bloch representation (EBR) \cite{asdb2014}. 

In brief, the possibilist transactional interpretation is a time-symmetric re-interpretation of standard quantum mechanics (SQM), originally proposed by John Cramer \cite{c1986} and then further elaborated by Ruth E. Kastner \cite{k2010,k2012,k2013}, adding to it the hypothesis that transactions do not form in our $3$-dimensional Euclidean space, but in a pre-empirical space (an idea that, as far as we know, is not supported by Cramer \cite{c2016}). More precisely, PTI (as well as Cramer's non-possibilist original version of the interpretation) transposes to the quantum domain the idea of Wheeler and Feynman of considering also advanced waves (travelling backward in time) in their interpretation of classical electrodynamics. This is done by observing that when taking the non-relativistic limit of Dirac or Klein-Gordon equations, one is left not with a single equation, but with two distinct equations: the usual Schroedinger equation, describing the evolution of Hilbert space ket-vectors, and its Hermitian conjugate, describing the evolution of the (dual) bra-linear forms. 

Contrary to SQM, both equations are considered to be physically relevant in PTI. More precisely, kets are associated with retarded offer waves of possibilities (OWP), evolving forward in time, which are emitted by those entities that, in a given context, play the role of sources (think of an atom in an exited state), whereas bras are associated to advanced confirmation waves of possibilities (CWP), evolving backward in time, emitted by quantum entities playing the role of absorbers, in response to the OWP received from a source. This a-temporal offer-confirmation dialogue between a source and its potential absorbers gives rise to a so-called \emph{incipient transaction}, which is the possibility for an \emph{actualized transaction}, a process that selects a specific outcome, in the form of an actual transfer of a given quantum of energy and momentum, detectable in our ordinary spatiotemporal theatre.

More precisely, let $S$ be a source encountering a given number $N$ of absorbers (detectors), labeled by the properties $A_i$, $i =1,\dots,N$. According to the transactional ontology, the retarded OWP, $|\psi\rangle$, emitted by $S$, stimulates the emission of advanced CWP, $\langle a_i|$, $i =1,\dots,N$, generated during the (potential) absorption of the OWP by the different detectors. This produces a feedback process between the source and the absorbers that, according to the transactional narrative, unfolds according to certain hypotheses. Forward and backward in time OWP and CWP evolutions between the source and the absorbers are assumed to produce equivalent effects (using a classical analogy of a wave actually propagating in space, one would say that the advanced wave travels across the same medium of the retarded wave, thus experiencing the same attenuation by it when propagating from the absorber to the source); the amplitudes of the CWP are assumed to be directly proportional to the incoming OWP (a proportionality precisely expressed by the scalar product); and $S$ is assumed to emit a possibly infinite collection of OWP, simultaneously, in accordance with the superposition principle and the fact that it is in the very nature of possibilities to be all available at once (every OWP emitted by $S$ is assumed to be echoed back from all absorbers, in accordance with the fact that in experiments a source is usually confronted not with a single absorber, but with a vast collection of equivalent absorbers, which can all simultaneously generate a convenient CWP). 

\section{Weighted symmetry breaking}

It is not the purpose of the present article to assess the plausibility of the above physical description and the associated assumptions. Certainly, until proven to the contrary, it is a possible description. However, we observe that it is insufficient to derive the Born rule. Indeed, the above account only describes a process leading to a weighted set of competing possible transactions, called incipient transactions. More precisely, these possible transactions can be associated with projection operators $|a_i\rangle \langle a_i |$ (obtained by matching the possible final OWP and CWP components), and the above ``dialogue'' between the source and the absorbers describes a deterministic transition from an initial pure state $|\psi \rangle \langle \psi |$ to a fully reduced mixed state:
\begin{equation}
|\psi \rangle \langle \psi |\to \sum_{j=1}^N |\langle a_j|\psi\rangle |^2 |a_j\rangle \langle a_j |.
\label{reduced}
\end{equation}
This, however, doesn't say how a potential transaction can become an actual transaction (and produce the so-called \emph{handshake}). In other terms, we have an explicit physical story in PTI, about how an incipient transaction forms, but there is no explicit physical story describing the realization of a particular transaction (outcome), as opposed to the many competing “incipient” ones. This is why in PTI an additional assumption is considered, which is the following: when there is an incipient transaction between a source $S$ and some absorbers $A_i$, $i =1,\dots,N$, as described by the above mixed state, the source-absorbers system can only select and actualize one of them, through a \emph{weighted symmetry breaking} (WSB) process; and the probability for the actualization of the $i$-th incipient transaction is precisely equal to $|\langle a_i|\psi\rangle |^2$, i.e., is given by the Born rule. 

More specifically, regarding the use of the notion of symmetry breaking in the PTI, Kastner writes \cite{k2013} (emphasis is ours): ``[...] spontaneous symmetry breaking can be consistently extended to PTI's account of the realized notion of one particular transaction out of several, or even many, incipient ones. The mathematics describing the situation provides us with a set of possible states of the system, but only one of those can be realized. The new feature appearing in the PTI account is that this set of possible outcomes is weighted by the square of the probability amplitude for that outcome. So, the proposed interpretation extends the basic principle of spontaneous symmetry breaking over a set of possible states: they certainly cannot all be realized (just as in the case of classical symmetry breaking), so \emph{the natural interpretation of the weight of a possible state is as a physical propensity, corresponding to an objective probability of the actualization of the state in question}. If we like, we can call this a ``weighted symmetry breaking'' or WSB.''

The problem in the above description is not the idea that potential transactions would be actualized by means of a symmetry breaking process, but the assertion that such symmetry breaking would be weighted by physical propensities precisely given by the square of the probability amplitudes. This is clearly a circular statement: if we want to explain the origin of the quantum mechanical Born rule, then we cannot use in our explanations the Born rule itself, even if we use the term `propensity' instead of `probability' to describe the statistical weights. 

We can see here a similarity between  PTI and \emph{decoherence theory} \cite{s2005}, where the transition (\ref{reduced}) is not described as the result of an interaction between OWP and CWP, but as the effect of an immersion of the physical entity in question in a heat bath, causing the off-diagonal elements of the initial state $|\psi \rangle \langle \psi |$, in the measurement's basis, to rapidly vanish, as the superposition gets ``diluted into the environment.'' But in decoherence theory, as well as in PTI, one still needs to explain how the pre-measurement state of the entity is ultimately collapsed, i.e., one has to explain why the different possible transitions
\begin{equation}
\sum_{j=1}^N |\langle a_j|\psi\rangle |^2 |a_j\rangle \langle a_j |\to |a_i\rangle \langle a_i|
\label{reduced2}
\end{equation}
happen with exactly the probabilities $|\langle a_j|\psi\rangle |^2$, $j=1,\dots,N$, when the same measurement is repeated many times. In other terms, one needs to explain, \emph{in a non-circular way}, why the experimental outcomes' relative frequencies are precisely the probabilities predicted by the Born rule. 

What we are emphasizing is that PTI, like decoherence theory, do not offer a possible solution to the measurement problem, as they are unable to derive the Born rule. They just provide an explanation for the transition of the initial vector (or ray) state to a pre-collapse fully reduced density matrix state. In decoherence theory the latter needs to be interpreted as a classical mixture of states, if one wants to establish a connection with the Born rule. This however would be an \emph{ad hoc} assumption, not defensible from the theory itself, which is why many adherents to decoherence theory are compelled to complete their scheme by resorting to a many-worlds picture (although this doesn't solve the problem of the derivation of the Born rule \cite{betal2000,b2007}). On the other hand, in PTI an ``incipient transaction state'' is not assumed to correspond to a classical mixture of actualized transactions, but to an objective condition of unstable equilibrium giving rise to a symmetry breaking process, finally producing a single outcome. Considering the PTI narrative, it is certainly natural, and in a sense plausible, to assume the existence of such a symmetry breaking, triggered by some uncontrollable fluctuations. However, what remains unexplained is why such process would happen exactly with the weights predicted by the Born rule. In other terms, if it is reasonable to assume, in the ambit of a realistic description of a quantum measurement, that the Born rule describes a physical process akin to a symmetry breaking, one still needs to explain why exactly the Born weights would result from it, and not others. 

\section{Hidden measurement-interactions}

As we are now going to explain, if PTI is unable to derive the Born rule, this is because it only describes the first element of a bipartite process, when measurements are non-degenerate, or of a tripartite process, when measurements are degenerate. But before doing so, we need to stress an important point. When considering a symmetry breaking process, what one is implicitly assuming is that \emph{hidden interactions} are at work. 
Let us explain in which sense they are hidden, using a simple example. Imagine a person pushing on a cylindrical stick vertically planted in the ground, along its vertical axis. By doing so, s/he can cause the stick to flex in an a priori unpredictable direction, breaking in this way the initial rotational symmetry and creating a direction that wasn’t present prior to the ``pushing experiment.'' When the stick flexes in a given direction, it does so because it is pushed in that direction. In other terms, this happens because a specific interaction between the hand and the stick is actualized, which produces a specific outcome. Of course, when the person vertically pushes on the stick, s/he cannot control all the infinitesimal fluctuations that will ultimately produce a hand-stick interaction slightly deviating from the vertical line. This means that prior to the pushing all these deviating hand-stick interactions are only potential, as only one will in the end be actualized, in a way that cannot be predicted in advance. For each flexing direction, we have a certain number of possible and equivalent hand-stick interactions that are all able to cause the stick to bend in that direction. Thus, if we have the same probability of obtaining the different possible directions, this means that we have the same number of hand-stick interactions that are available to be actualized for each of them. 

So, as the stick example illustrates, a symmetry breaking process is a process of actualization of a hidden interaction. Prior to the process, these interactions remain `hidden' as they are not yet actual interactions and one cannot predict in advance which one will be actualized, the process of actualization being not under the control of the experimenter (here the person pushing on the stick). As we will explain, the interactions possibly intervening in a quantum measurement are hidden precisely for these reasons, and in fact also for an additional one, but for the moment let us reflect on a key aspect of this description. In the stick example, as we said, one expects to obtain the same probability for each direction. However, it is easy to imagine experimental situations where there would be a bias towards certain directions; these are situations that we could conveniently call, in accordance with Kastner, of WSB. Situations of this kind could happen because the stick would not be planted in the ground in a perfectly vertical way, or because the `way the person pushes onto it' may favor certain directions in a systematic way. But whatever the reasons for having certain directions being selected more frequently than others, we can certainly explain these differences by considering that the number of hidden hand-stick interactions available to be actualized is not anymore the same for each direction, and that the observed experimental probabilities precisely express this difference in the relative number of potential interactions associated with each outcome-direction. 

If we want to describe the quantum probabilities as a WSB process, as suggested by Kastner, we thus need to show that the Born rule can be derived as a process governed by hidden-measurement interactions, i.e., as a process where a predetermined number of hidden interactions can be associated to each possible outcome, so that the relative numbers of these interactions precisely correspond to the Born rule of probabilistic assignment. What is interesting, and remarkable, is that such hidden-interaction mechanism is actually built in the quantum formalism, which means that when we derive the Born rule using the HMI and the associated EBR, we  follow a methodological route very similar to that of the PTI. Indeed, in the same way as the PTI does not posit new mathematical structures, but simply postulate a physical referent for those mathematical entities that (according to the PTI perspective) remained uninterpreted in the standard formalism (the advanced states appearing in the Hilbert space inner products, in the complex conjugate Schroedinger equation, and in the Born Rule), in the same way the hidden measurement program is implemented by interpreting the density matrix states also as a description of pure states (and not just as classical statistical mixtures of states), and the hyper-surfaces generated by the outcome-states, within the generalized Bloch representation, as a description of the hidden measurement interactions available in that specific measurement context. 

\section{Extended Bloch representation}

We are now going to describe the unfolding of a measurement process according to the HMI. We will do so by using the specific terminology of the PTI. This of course is not necessary, and alternative narratives can be also considered. What however needs to be part of all possible narratives, is the idea that there are elements of reality (or `beables', to use Bell's designation) describable as `hidden measurement-interactions', which are responsible for producing one of the outcomes (\ref{reduced2}), in accordance with the Born rule. We start by explaining how states can generally be represented in the generalized Bloch sphere. 

Let $D$ be a density matrix in a $N$-dimensional Hilbert space ${\cal H}$. We introduce a determination of the generators of $SU(N)$ (the special unitary group of degree $N$), which are traceless and orthogonal self-adjoint matrices $\Lambda_i$, $i=1,\dots, N^2-1$, that can be chosen to be normalized as ${\rm Tr}\, \Lambda_i\Lambda_j=2\delta_{ij}$. Then, one can show that there is a unique $(N^2-1)$-dimensional real vector ${\bf r}$ such that $D\equiv D({\bf r})$ can be written \cite{asdb2014}:
\begin{equation}
D({\bf r}) = {1\over N}\left(\mathbb{I} +c_N\, {\bf r}\cdot\mbox{\boldmath$\Lambda$}\right) = {1\over N}\left(\mathbb{I} + c_N\sum_{i=1}^{N^2-1} r_i \, \Lambda_i\right),
\label{formulaNxN}
\end{equation}
where we have defined $c_N\equiv \sqrt{N(N-1)/2}$. In other terms, we can represent states by means of real vectors in a generalized unit Bloch sphere $B_1(\real^{N^2-1})$. For $N=2$, we have only 3 generators and they correspond to the well known Pauli matrices, whereas for $N=3$ we have 8 generators that can be chosen to correspond to the so-called Gell-Mann matrices. The $N=2$ case is special, as the $3$-dimensional vectors ${\bf r}$ representing the states completely fill the $3$-dimensional unit Bloch sphere. Instead, for $N\geq 3$, the $(N^2-1)$-dimensional generalized Bloch sphere is not completely filled with states, but contains a convex region filled with states, inscribed in the latter, whose shape and orientation depend on the specific choice of generators. Also, it is easy to show that ray-states (one-dimensional projection operators, or pure states, in the standard quantum terminology) are always represented by unit vectors ($\|{\bf r}\|=1$), whereas density matrix states are described by vectors that are internal to the sphere ($\|{\bf r}\|\leq1$). 

We will not enter into the mathematical details of this generalized Bloch representation, for which we refer the interested reader to \cite{asdb2014} and the references cited therein. What is important to mention here, for the purpose of our discussion, is that the if $\{|a_1\rangle,\dots,|a_N\rangle\}$ is a basis of ${\cal H}$, for instance the basis associated with the spectral family of a given observable, then to each projection $ |a_i\rangle \langle a_i |$ we can associate a $(N^2-1)$-dimensional unit vectors ${\bf{n}}_{i}$ in the generalized Bloch sphere, such that $ |a_i\rangle \langle a_i | = {1\over N}\left(\mathbb{I} +c_N\, {\bf n}_i\cdot\mbox{\boldmath$\Lambda$}\right)$, and it is possible to show that ${\bf n}_i\cdot {\bf n}_j = -{1\over N-1}$ for $i\neq j$. This means that 
the $N$ vectors ${\bf{n}}_{i}$, representative of the outcome-states $|a_i\rangle \langle a_i |$, form a $(N-1)$-dimensional simplex $\triangle_{N-1}$, inscribed in the generalized $(N^2-1)$-dimensional Bloch sphere $B_1(\real^{N^2-1})$. 

\begin{figure}[!ht]
\centering
\includegraphics[scale =.18]{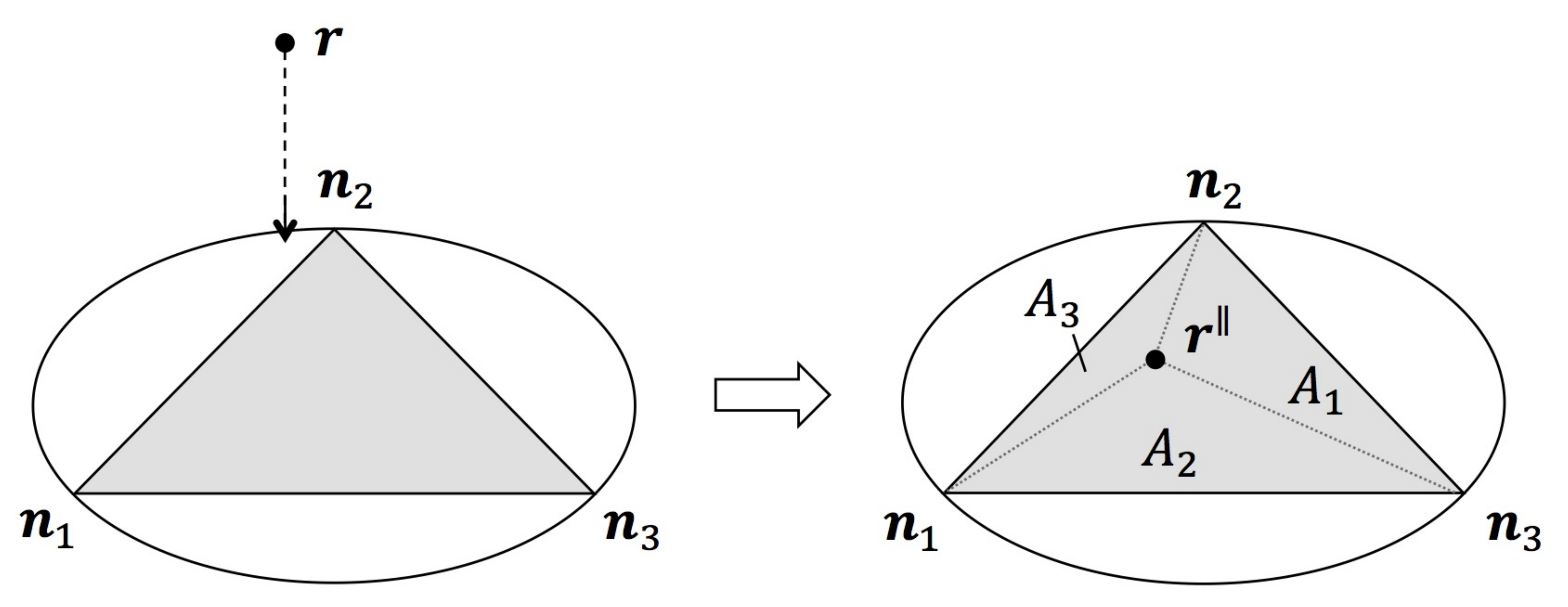}
\caption{The formation of an incipient transaction, from the viewpoint of the generalized Bloch sphere. The three unit vectors ${\bf{n}}_{1}$, ${\bf{n}}_{2}$ and ${\bf{n}}_{3}$, representative of the outcome states, determine a $2$-dimensional equilateral triangle. The abstract point particle representative of the initial state $|\psi \rangle \langle \psi |$, initially located at ${\bf{r}}$ (at the surface of the $8$-dimensional Bloch sphere, which of course cannot be represented in the drawing),  transition to the on-triangle position ${\bf r}^\parallel$, representative of the fully reduced state $\sum_{i=1}^N |\langle a_i|\psi\rangle |^2 |a_i\rangle \langle a_i |$, by following a path orthogonal to the triangle's plane. This generates three disjoint convex regions $A_{1}$, $A_{2}$, and $A_{3}$, representative of the hidden measurement-interactions that are available to actualize the three possible outcomes ${\bf{n}}_{1}$, ${\bf{n}}_{2}$ and ${\bf{n}}_{3}$, respectively.
\label{triangolo}}
\end{figure}

For simplicity, we just consider here the non-degenerate situation with three outcomes $(N=3$), the general situation being a straightforward generalization of it. According to the PTI narrative, incipient transaction are established through OWP-CWP encounters in a pre-spatiotemporal (PST) realm, which here we can identify with the $8$-dimensional generalized Bloch sphere in which the simplex generated by the three outcome-states (an equilateral triangle) is inscribed. It can then be shown that the establishment of an incipient transaction, corresponding to transition (\ref{reduced}), corresponds to a movement of the abstract point particle representative of the initial state $D({\bf r})= |\psi\rangle \langle \psi |$, from position ${\bf r}$ at the surface of the sphere, to position ${\bf r}^\parallel$ on the $2$-dimensional simplex $\triangle_{2}$, along a path orthogonal to the latter (see Fig.~\ref{triangolo}), with ${\bf r}^\parallel$ the vector representative of the fully reduced mixed state, i.e., $D({\bf r}^\parallel)=\sum_{j=1}^N |\langle a_j|\psi\rangle |^2 |a_j\rangle \langle a_j |$.

In other terms, the (deterministic) establishment of an incipient transaction corresponds to the abstract point particle plunging into the sphere and reaching a specific position on the measurement simplex. At this point, we can observe that the on-simplex vector ${\bf r}^\parallel$ gives rise to a partitioning of $\triangle_{2}$ into three distinct triangular subregions $A_1$, $A_2$ and $A_3$, as illustrated in Fig.~\ref{triangolo}. Subregion $A_1$ has vertexes ${\bf n}_{2}$, ${\bf n}_{3}$ and ${\bf r}^\parallel$, subregion $A_2$ has vertexes ${\bf n}_{1}$, ${\bf n}_{3}$ and ${\bf r}^\parallel$, and subregion $A_3$ has vertexes ${\bf n}_{1}$, ${\bf n}_{1}$ and ${\bf r}^\parallel$. According to the EBR, these subregions of $\triangle_{2}$ are precisely the referents for the hidden interactions responsible for the indeterministic collapse of the reduced state into one of the available outcomes. More precisely, we can consider each point ${\mbox{\boldmath$\lambda$}}\in \triangle_{2}$ as the representative of a potential interaction, able to deterministically produce an outcome. Which of these interactions is associated with which outcome precisely depends on the partitioning produced by the abstract point particle. More precisely, all the hidden interactions belonging to $A_i$ are responsible for the outcome $|a_i\rangle \langle a_i |$, represented by the unit vector ${\bf n}_{i}$, $i=1,2,3$. In other terms, when considering the structure of the generalized Bloch sphere, inherited by the geometry of the Hilbert space, we see that its interpretation can be extended to provide an explicit description of the WSB mechanism that was only postulated (but not explained) in the PTI. 

It is important to emphasize that we are not resorting here to some \emph{ad hoc} mathematical structures to describe the indeterministic collapse, which would be absent from quantum theory: we are just interpreting structures already appearing in the theory, which traditionally have not received any specific interpretation. This way of proceeding should be particularly appealing to the proponents of TI and PTI, as this is precisely how they also proceeded, when postulating a physical referent for the advanced states appearing in the Hilbert space inner products. Now, is our interpretation of the points in the simplex, as referents for the interactions producing the different outcomes, sufficient to derive the Born rule, as a WSB process? The answer is affirmative. Indeed, if all points in $\triangle_{2}$ correspond to potential interactions that can be equally actualized during the mesurement, and if those belonging to subregion $A_i$ are precisely those responsible for the outcome ${\bf n}_{i}$, then the probability $P({\bf r}^\parallel\to{\bf n}_{i})$ for the transition (\ref{reduced2}) has to be given by the ratio: 
\begin{equation}
P({\bf r}^\parallel\to{\bf n}_{i})= {\mu (A_i)\over \mu (\triangle_{2})}, \quad i=1,2,3,
\label{Born}
\end{equation}
where $\mu$ denotes the Lebesgue measure, and $\mu (\triangle_{2})= 3\sqrt{3}/4$. By a straightforward calculation, it is then possible to show that the r.h.s. of (\ref{Born}) is exactly equal to $|\langle a_i|\psi\rangle|^2 = {\rm Tr}\, D({\bf r}^\parallel) |a_i\rangle \langle a_i| = {\rm Tr}\, D({\bf r}) |a_i\rangle \langle a_i|$, and that this remains true in the general situation of a measurement having an arbitrary number $N$ of outcomes \cite{asdb2014}. In other terms, the EBR of quantum mechanics provides a natural explanation of quantum measurements as processes resulting from situations of unstable equilibrium, where hidden measurement-interactions mutually compete in the actualization a specific outcome. Which one of the available interactions will each time be actualized depends of course on the fluctuations that are inevitably part of the measurement context, but the more interactions are available to produce a given outcome and the more probable such outcome will be. This is precisely expressed by the ratio (\ref{Born}), which is perfectly equivalent to the Born rule, and this is also the missing description of the WSB process postulated in the PTI.

\begin{figure}[!ht]
\centering
\includegraphics[scale =.18]{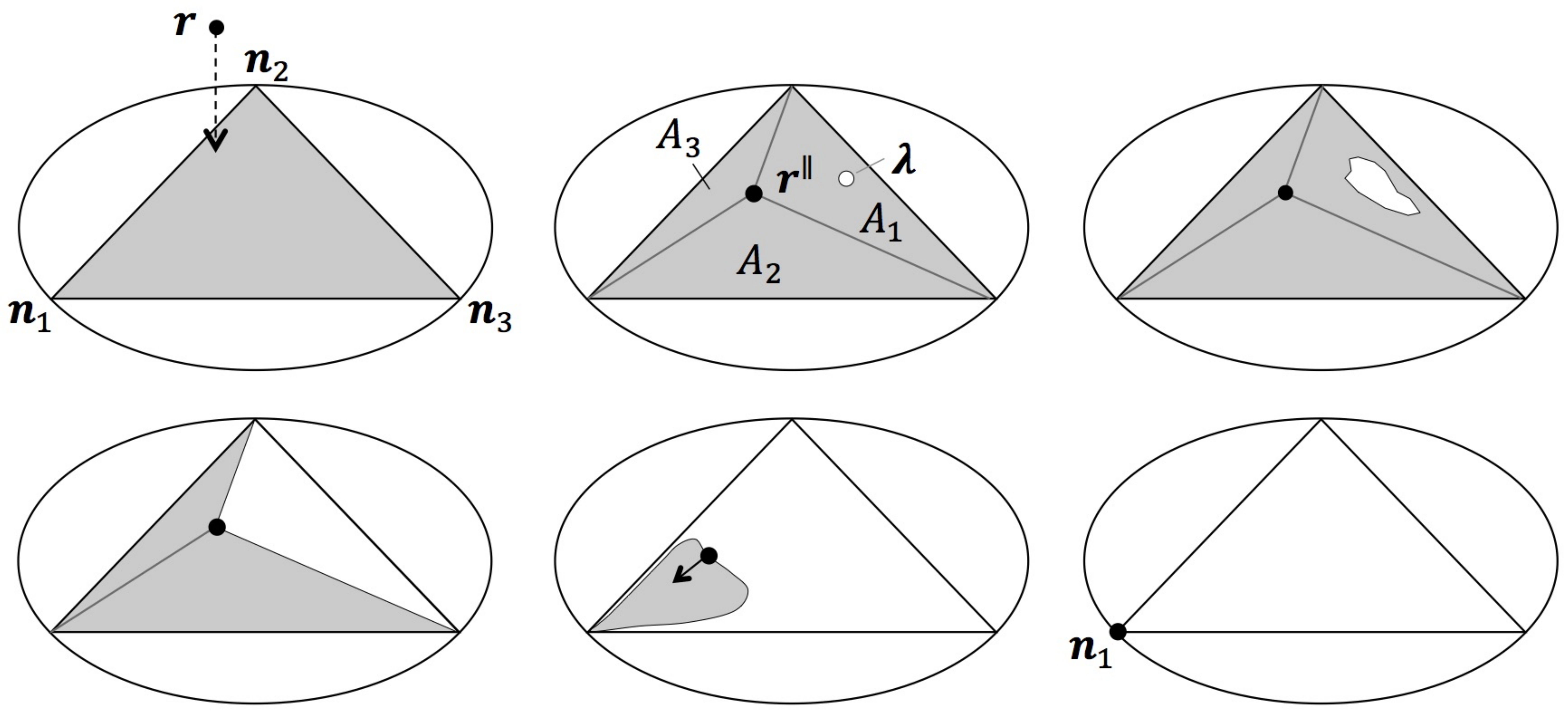}
\caption{The unfolding of a measurement process with three distinguishable outcomes: ${\bf n}_1$, ${\bf n}_2$ and ${\bf n}_3$. The abstract point particle representative of the state, initially located in ${\bf r}$, deterministically approaches the triangular elastic membrane along an orthogonal path, reaching the on-membrane point ${\bf r}^\parallel$, defining in this way three subregions $A_1$, $A_2$ and $A_3$. The abstract elastic membrane then disintegrates at some unpredictable point \mbox{\boldmath$\lambda$}, here assumed to be in subregion $A_1$, causing it to lose two of its three anchor points and drawing in this way the point particle to its final location, here ${\bf{n}}_1$.
\label{collapse}}
\end{figure}

A few remarks are in order. First of all, we refer the interested reader to \cite{asdb2014,asdb2015a,asdb2015b}, to take due knowledge of all the mathematical aspects of the EBR. In these references, a mechanical-like picture is also given, where the measurement simplex is described as if it was an elastic and disintegrable hyper-membrane, which by contracting is able to bring the abstract point particle to one of the vertices of the simplex, in accordance with the projection postulate. More precisely, one can think of the potentiality region of the simplex as a uniform elastic substance which can  attract the point particle, causing it to stick onto it, at point ${\bf r}^\parallel$. The instability of the substance will then produce its disintegration, starting from an unpredictable point \mbox{\boldmath$\lambda$}, belonging to one of the three subregions $A_1$, $A_2$ or $A_3$, delineated by the line segments connecting the particle's position to the vertex points. In this description one has to think of these line segments as ``tension lines'' altering the functioning of the membrane, in the sense of making it less easy to disintegrate along them. The disintegration can then only propagate inside the specific subregion where it starts, causing the two associated anchor points to tear away, producing the detachment of the membrane, which being elastic then contracts towards the only remaining anchor point, drawing to that position also the point particle attached to it, corresponding to the outcome of the measurement (see Fig.~\ref{collapse}). 

Considering the PTI account, one can of course interpret this abstract membrane's collapse as a description of the absorption of the entity's OWP by one of the $N$ available detectors. But of course, the details of how this process truly unfolds, behind the spatiotemporal scenes, to produce an actualized transaction, or more simply an outcome state, are not so important. What is important is that the HMI provides a very general mechanism for the description of a measurement, which can be used to derive the Born rule when the set of states is Hilbertian (or even more general rules of probabilistic assignment when the set of states is non-Hilbertian; see for instance \cite{asdb2014b,asdb2015d,asdb2015e,asdb2015f}) and  is able to explain the indeterministic collapse of a fully reduced density matrix state to an outcome state as an objective WSB process. 

The description of more general measurement situations, when certain outcomes are experimentally undistinguishable, is also possible within the HMI and the associated EBR. For this, one simply has to ``fuse'' together the subregions associated with the same degenerate outcome. This will alter the collapse of the membrane and the point particle will not anymore necessarily be drawn to a vertex point of the simplex, but can also be drawn to one of its sub-simplexes. A final purification-like process then needs to be considered in this case, to bring the collapsed state back to the surface of the sphere, in accordance with the the L\"uders-von Neumann projection formula (see \cite{asdb2014} for the details).

It is interesting to note that in the same way an incipient transaction is considered in PTI to be a real physical condition of interaction between a source and the absorbers, in the EBR a density matrix state is also considered to be a a condition truly describing the reality of the measured entity during the measurement process. Therefore, both the PTI and the EBR accounts suggest that the notion of pure state (intended here as a notion describing an objective, and not subjective, condition of the entity under consideration) should be extended to also include density matrices. By the way, the latter would not only be the missing states without which it is impossible to properly describe a measurement, but also those without which it is impossible to understand, without paradoxes, the very notion of entanglement (see the analysis in \cite{asdb2016}). 

\section{Non-spatiality}

We have previously mentioned that the hidden interactions responsible for a WSB process in a quantum measurement are ``more hidden'' than those intervening in an experiment like the one with the cylindrical stick planted in the ground. This is so because these interactions are not only potentials, at the start of the measurement, but are also \emph{non-spatial}, or more generally non-spatiotemporal, exactly as in the PTI account incipient transactions are also considered to be dynamical events taking place in a pre-spatiotemporal (PST) realm. In that respect, it is worth observing that the passage from Cramer's ``standard'' TI to Kastner's ``extended'' PTI was operated by the latter author because of difficulties the TI narrative faced, like the interpretation of multiparticle offer and confirmation waves, the possibility of causal loop leading either to inconsistency or inconclusive quantitative predictions (see \cite{Maudlin2002}, pp- 199-200, and \cite{Berkovitz2002,Berkovitz2008}), and of course the nature of the WSB process leading to an actualized transaction \cite{k2013}. To solve these difficulties Kastner proposed that offer and confirmation waves should not be considered as spatial waves living in the three-dimensional Euclidean space, but as dynamically efficacious possibilities forming a higher-dimensional space, indicated as ``pre-spacetime'' (see however \cite{c2016}, for Cramer's new approach regarding these dimensionality questions, which according to him would make the retreat to higher-dimensions unnecessary).
 
It is important to emphasize that the hypothesis of a non-spatial realm describing the reality of quantum entities and their interactions is a inescapable hypothesis if one considers that: (1) quantum measurements are not just discovery processes, i.e., observations of properties that are already actual prior to the observation, but also, in part, creation processes, provoking a real change of the measured entities; (2) the states of quantum entities are descriptions of their objective modes of being. This means that before a measurement of, say, the position of an electron prepared in a superposition state (with resepct to the position observable), one cannot affirm that the electron was in some unknown place prior to its localization. As a consequence, an electron (and in general all microscopic physical entities) are to be considered fundamentally non-spatial (and only in that sense non-local) entities. This means that the wave function describing an electron cannot be interpreted as a physical wave that would be present in space, but as a description of the  availability of the electron in being spatialized in the different possible regions (with the Born rule predicting with certainty the different localization probabilities, which therefore are to be interpreted as elements of reality in the Einstein-Podolsky-Rosen sense). And the same also holds of course for the other potential properties of a microscopic quantum entity, like (linear and angular) momentum. Again, an electron will generally not possess a well-defined momentum before its measurement, as the latter will only be partly created by the measurement itself. 

What we are stressing here is that the notion of \emph{non-spatiality}, i.e., of a pre-empirical PST layer of our reality, as Kastner calls it, is in fact implicit in any realistic approach to quantum mechanics that would take seriously its formalism (see for instance the discussion in \cite{asdb2015a}). This means that the fourth assumption mentioned in the beginning of this note actually follows from the first two, and this is the reason why the key notion of non-spatiality was introduced by one of us as early as the late eighties \cite{a1990} and discussed since then in a number of works, in what has been called the \emph{creation-discovery view}; see \cite{a1998, a1999} and the references cited therein, and the discussions in \cite{sdb2011,sdb2012,sdb2013}.

\section{Conclusion}

To conclude, our attempt in the present note was to frame part of the PTI ontology in the ambit of the HMI ontology, of the associated creation-discovery view and EBR. Regarding the latter, for what concerns the description of the initial preparatory stage of a measurement, during which the abstract point particle plunges into the generalized Bloch sphere to reach the measurement simplex, we emphasize that the EBR does not prescribe any specific physical mechanism for producing such transition. Therefore, one can adopt here the PTI narrative, or that of decoherence theory, or any other physical account able to explain how a vector state, in a measurement context, can deterministically transform into a fully reduced mixed state, with the latter being interpreted not as a mixture, but as the real pre-collapse condition of the measured entity in that specific experimental context. But then, a WSB will have to take place, and for its description one cannot avoid introducing the key notion of hidden measurement-interactions, if one  wants to explain not only qualitatively but also quantitatively the emergence of the quantum (Born) weights. Again, the EBR is agnostic regarding the physical origin of these hidden interactions: all we know for the time being is that they naturally emerge from the quantum formalism, when state vectors are ``squared'' and represented in the generalized Bloch sphere. 

The HMI certainly reinforce the idea that quantum measurements should be interpreted as WSB processes, similarly to experimental situations  occurring in classical physics, like in our example of the flexing of a stick. Indeed, the quantum behavior of an entity would not be so much a consequence of its microscopic or macroscopic nature, but of the way one decides to actively experiment with it, by means of specific protocols. This is why it is possible to replicate all sorts of quantum effects also by acting on macroscopic entities working at room temperature \cite{a1986,a1998,a1990,a1999,sdb2013}, and that the quantum formalism has been so successfully applied in the past decade to also model human cognition processes (see \cite{asdb2015d,asdb2015e} and the references cited therein). But, as we said already, the dramatic difference between microscopic and macroscopic entities is that only the latter can be attached a permanent presence in our spatiotemporal theater, whereas the former are only ``dragged'' into it because of their interaction with a detection system or, to use the PTI jargon, because a pre-spatiotemporal incipient transaction is actualized, giving rise to an observable phenomena in the laboratory.

\end{document}